# Rotor-phonon coupling in perovskite CH₃NH₃PbI₃: the origin of exceptional transport properties

Bing Li[1], Yukinobu Kawakita[1], Yucheng Liu[2], Masato Matsuura[3], Kaoru Shibata[1], Seiko Kawamura[1], Takeshi Yamada[2], Kenji Nakajima[1] & Shengzhong (Frank) Liu[2,4]

[1]Japan Proton Accelerator Research Complex, Japan Atomic Energy Agency, Tokai, Ibaraki 319-1195, Japan.

[2]Key Laboratory of Applied Surface and Colloid Chemistry, National Ministry of Education, Institute for Advanced Energy Materials, School of Materials Science and Engineering, Shaanxi Normal University, Xi'an 710119, P. R. China.

[3]Neutron Science and Technology Center, Comprehensive Research Organization for Science and Society, Tokai, Ibaraki 319-1106, Japan.

[4]Dalian Institute of Chemical Physics, Dalian National Laboratory for Clean Energy, Chinese Academy of Sciences, Dalian 116023, P. R. China.

Correspondence and request for materials should be addressed to B. L. (bing.li@j-parc.jp) or S. L. (liusz@snnu.edu.cn)

**Atomic dynamics takes a fundamental part in numbers of physical properties of solids like high-$T_c$ superconductivity[1], semiconducting transports[2], and thermoelectricity[3]. Perovskite CH₃NH₃PbI₃ exhibits outstanding photovoltaic performances, but the exact physical scenario has not been established yet, due to the inadequate understanding of the atomic dynamics and exceptional transport properties[4-7]. We present a complete atomic dynamic picture consisting of phonons, rotational modes of protons and molecular vibrational modes, which is constructed by carrying out high-resolution time-of-flight inelastic neutron scattering measurements in a wide energy window ranging from 0.0036 to 54 meV on a large single crystal sample. A three-fold rotational mode of protons activated around 80 K reduces the lifetimes of acoustic and optical phonons down to about 4.5 ps and below 1 ps at 150 K, respectively. The orthorhombic to tetragonal phase transition takes place with a slower four-fold rotational mode of the C-N axis concomitantly setting in at ~ 165 K, above which the optical phonons are too broadened to be distinguished whereas the acoustic ones are still robust. The significantly reduced lifetimes of optical phonons are linked to the smaller mobility of charge carriers while the ultralow lattice thermal conductivity is attributed to nanoscale mean free paths of acoustic phonons. These microscopic insights provide a solid standing point, on which perovskite solar cells can be understood more accurately and their performances are perhaps further optimized. The** **revealed rotor-phonon coupling opens up an emergent opportunity to create unprecedented functionalities of materials.**

Over last few years, the inorganic-organic hybrid perovskite solar cells have taken the worldwide research community by storm[4]. As a representative, the energy conversion efficiencies of $CH_3NH_3PbI_3$-based solar cell have been improved from the starting 4% in 2009 to higher than 22% in 2015[8,9]. Very recently, the successes in growth of inch-sized high-quality single crystals and the device integration on wafers have paved the route to large-scale practical photovoltaic applications[10,11]. In spite of explosive progresses on $CH_3NH_3PbI_3$, transport behaviours, as one of core physical properties, have not been well-understood yet. The mobility of charge carriers is relatively smaller compared with classical semiconductor GaAs that has a similar band gap[4-6,12,13] and the scattering mechanism is still under debate[14,15]. Resembling the charge transport, the thermal transport is unusual as well. The thermal conductivity is surprisingly low, about 0.5 $Wm^{-1}K^{-1}$ for single crystals at room temperature[7], which is tenfold lower than that of GaAs[13] and is even lower than that of amorphous silicon[16]. These peculiarities apparently differing from conventional photovoltaic semiconductors suggest that there is a different scenario dominant in $CH_3NH_3PbI_3$. However, the atomic-level description of $CH_3NH_3PbI_3$ is complicated by the *hybrid* nature where *hard* inorganic crystalline frame and *soft* organic cation coexist and may also interact via hydrogen bonds between H and I[17-19]. In order to achieve a complete understanding of $CH_3NH_3PbI_3$, consequently, both local molecular motions and phonon excitations have to be taken into account.

The best approach for this issue is, without a doubt, inelastic neutron scattering. The lattice dynamics calculations indicate that low-energy phonons are entirely composed of motions of Pb and I[17,18], which are easily excited at relatively lower temperatures and hence take main parts in determining the physical properties. These phonons can be efficiently probed throughout the Brillouin zones, due to the larger coherent scattering cross-sections of Pb and I. Simultaneously, the incoherent scattering cross-section of H assures the individual motions of protons can be determined by extracting the quasi-elastic broadening underneath the elastic line. We combine two time-of-flight neutron spectrometers to search the dynamics in a wide energy window ranging from 0.0036 to 50 meV. The high-resolution is beneficial to minimizing the contamination of phonons by the incoherent scattering of H. A large single crystal weighted about 2.23 grams is used to enhance the statistics of measurements. We observe phonons, rotational modes, and local vibrational modes across the low-temperature phase transition.

These microscopic results suggest the phonons are intimately coupled with rotational modes and their interplay dominates the unique properties of this system.

$CH_3NH_3PbI_3$ crystallizes in the common perovskite structure where the organic cation $[CH_3NH_3]^+$ occupies the centre of the $PbI_6$ octahedral cage, as shown in **Fig. 1a**. The organic cation is a robust rotor at finite temperatures, which leads to dynamic disorders. It is also linked to the octahedra through the strong hydrogen bond to $I$[16,17]. Intensive efforts have concentrated on the structural analysis and a census turns out to be reached. With decreasing temperature, it undergoes successive phase transitions, from cubic ($Pm\bar{3}m$) to tetragonal ($I4/mcm$) at 330 K, and then to orthorhombic ($Pnma$) at 165 K[20]. Our single crystal exhibits a sharp phase transition between 160 and 165 K, as shown in **Fig. S1**, which is in agreement with previous reports. The reciprocal space is mapped out along $[hh0]$ and $[00l]$ directions by using the elastic neutron scattering measurements on the crystal. As compared in **Fig. 1 c** and **d**, the phase transition to the orthorhombic phase is characteristic of the presence of superlattice reflection $\left(\frac{3}{2}\frac{3}{2}1\right)$.

Current reports on rotational dynamics are contradictory[21,22], probably owing to the lack of results on single crystals. Given that this compound is extremely sensitive to ambient conditions[23], measurements on a single crystal (with minimized surface contribution) are anticipated to deliver more intrinsic information. The accurate rotational dynamics of protons is obtained by performing ultrahigh energy resolution quasi-elastic neutron scattering (QENS) measurements on the large single crystal at the backscattering neutron spectrometer, DNA. Firstly, we demonstrate the rotational dynamics in the orthorhombic phase. As shown in **Fig. 2a**, the spectrum at 50 K is basically described by a delta function convoluted to the instrumental resolution (the full width at half maximum is around 0.0036 meV). Around 80 K, the QENS signal described by a Lorentzian function is identified underneath the elastic line, indicative of the activation of rotations. The QENS intensity becomes more noticeable at 140 K. The momentum-transfer averaged spectra along $[00l]$ direction are systematically fitted by including a Lorentzian function, a delta function, and a constant background, which are convoluted to the instrumental resolution. An example is given at 140 K in **Fig. 2b**. The half width at half maximum of the Lorentzian function, $\Gamma$, is examined as a function of momentum transfer and temperature. As shown in **Fig. 2e**, $\Gamma$ is almost independent on $l$. Since $\Gamma$ is inversely related to the relaxation time, such a value gives rise to an average relaxation time of $\sim 23(1)$ ps at 140 K. The temperature dependence is fitted to the Arrhenius relation and the activation energy about 47.9(6) meV per chemical formula is obtained (**Fig. 2f**). The elastic incoherent structure factor (EISF) is shown in **Fig. 2g** at 140 K, well described by the three-fold rotation ($C_3$) model. In this model, $EISF_{C_3} = \frac{1}{3}\left[1 + 2j_0\left(Qd_{H-H}\right)\right]$[24], where $j_0$ is zeroth-order spherical Bessel function and $d_{\text{H-H}}$ is the average distance between two adjacent protons, about 1.72 Å in terms of neutron powder diffraction results[18] (see **Fig. S3** for details).

In the tetragonal phase, the rotational dynamics becomes more complicated. The spectrum obtained at 180 K at DNA displays a narrower QENS than that at 140 K, as compared in **Fig. 2b** and **c**. The relaxation time is determined to be 64(2) ps. This is unusual, because rotational dynamics is thermally activated. To understand this mode, likewise, we consider a rotational model. Shown in **Fig. 2h** is the comparison between the experimental EISF with that of the four-fold rotational ($C_4$) model, where $EISF_{C_4} = \frac{1}{4}\left[1 + 2j_0\left(Qd_{H-H}\right) + j_0\left(\sqrt{2}Qd_{H-H}\right)\right]$[24] and $d_{\text{H-H}}$ is 2.18 Å[18] (**Fig. S3**). It can be seen that the experimental data is well reproduced by this model, indicating that a four-fold rotational mode sets in after the phase transition to tetragonal phase. At 180 K, the $C_3$ mode is found by using the cold neutron time-of-flight spectrometer AMATERAS, which provides a much wider energy window. As can be seen in **Fig. 2d**, a broad Lorentzian function is found, corresponding to a relaxation time as short as 0.71(3) ps. Hereto, we are able to conclude on the rotational dynamics of protons. In orthorhombic phase, the protons in $CH_3$ and $NH_3$ undergo jumping rotations with respect to the C-N axis. The relaxation time is dramatically shortened in the tetragonal phase while the C-N axis starts to rotate with respect to the $c$ axis of the crystal structure. The rotational modes are illustrated in **Fig. 1B**. Our results about the proton rotational dynamics are in agreement with the recent QENS measurements on a powder sample[22].

The local dynamics is well understood above and now we move to the collective dynamics, i.e., phonons, which are measured by choosing several incident energy of neutrons ($E_i$) at AMATERAS. The obtained dynamic structure factor $S(\boldsymbol{q},E)$ is shown in **Fig. 3** at 5 K in the direction of [00$l$]. With $E_i$ of 54 meV, the full view of $S(\boldsymbol{q},E)$ is provided and it can be seen there are four intense bands present. The momentum-transfer averaged spectrum at $10 < l < 14$ (**Fig. 3d**) indicates these four modes are located at 11.2, 15.1, 24.3, and 37.6 meV, respectively. With smaller $E_i$, more details are revealed. We find the peak at 11.2 meV shown in **Fig. 3a** and **d** is actually a superposition of three peaks at 10.7, 11.5, and 12.6 meV with $E_i = 16$ meV. The data with $E_i$ of 7 meV suggests four low-energy modes at 2.28, 3.11, 3.79 and 4.35 meV. According to the recent lattice dynamics calculations[14,15], the mode at 2.28 meV is assigned as the acoustic phonons at the zone boundaries decorated by strong optical phonons. Other flat modes, except two at 11.5 and 37.6 meV, might be optical phonons. The intensity of modes at 11.5 and 37.6 meV are almost momentum transfer independent and much stronger than other modes. Raman scattering studies indicate that the former may be related to the lurching mode of the molecules and the latter is associated with the torsion of C-N axis[18]. Indeed, these assignments are experimentally supported by the inelastic neutron scattering study on $CH_3NH_3PbBr_3$ and its deuterated reference[25]. In this comparison, a low-energy mode around 5 meV is present in both cases, while an intense mode at approximately 12 meV disappears in the deuterated sample with two peaks besides remaining. Thus, the mode at 5 meV is attributed to phonons while that at 12 meV is related to incoherent scattering cross section of H involved in molecular vibrations. The phonon energy (5 meV) is higher than that (2.28 meV) of $CH_3NH_3PbI_3$, probably because Br is much lighter than I.

Detailed acoustic phonons of the Brillouin zone (220) are examined with $E_i$ of 4 meV. Shown in **Fig. 4a-d** are the longitudinal and transverse phonons at 5 and 180 K, respectively. Even though the background level is high due to the huge incoherent scattering cross section of H, the dispersions can be seen in both directions. At 5 K, the dispersion curve of longitudinal acoustic (LA) phonon emanates from zone center of (220), undergoes a steep increase and finally approaches the top about 2.28 meV in the middle of zone centres and zone boundary ($h \sim 0.2$), where it is overlapped with longitudinal optical (LO) phonons. The dispersion of transverse acoustic (TA) phonon exhibits a moderate slope. Intense transverse optical (TO) phonons are also present. At 180 K, the intensity of phonons is fairly enhanced and we determine the dispersions of LA and TA phonons by fitting the momentum-transfer (filled circle) or energy-transfer (filled square) averaged spectra. The solid lines represent the slopes of dispersion curves near the zone centers, which determine the velocities: $v_{LA}$ = 2841 ms[-1] and $v_{TA}$ = 1155 ms[-1]. Shown in **Fig. 4e** is the fitting of the TA phonon, which derives the lifetime of 3.61(42) ps in the middle of the zone boundary and zone center. For the LA case, we determine that the lifetime is about 4.39(46) ps at 180 K in the vicinity of the zone center (see **Fig. S5**). Nevertheless, the LA phonon collapses near the zone boundary along [$hh0$], which suggests a zone-boundary transition in concert with the absence of the superlattice $\left( \frac{3}{2} \frac{3}{2} 1 \right)$[26].

More strikingly, the LO and TO phonons around 2.28 meV completely vanish at 180 K. The data of the (004) zone (**Fig. S6**) indicates that they survive at 150 K. Thus, we compare the optical phonons located at 10.7 and 12.6 meV at 5, 150 and 180 K, which are two strongest optical phonons observed in the spectra. At 5 K, these optical phonons are well-defined, as shown in **Fig. 3e**. With warming up to 150 K, as shown in **Fig. 4f**, the lifetimes of TO phonons (upper panel) at 10.7 and 12.6 meV are reduced down to 0.91(15) and 0.62(4) ps while those of LO phonons (lower panel) are reduced down to 0.82(5) and 0.57(2) ps, respectively. These lifetimes are much shorter than 4.53(9) ps for the TA phonon (inset of **Fig. 4e**). At 180 K, just above the phase transition, both TO and LO phonons are too broadened to be distinguished. This means the actual lifetimes at 180 K are much shorter than those at 150 K. Note that the peak in between these two optical phonons is associated with the molecular vibrational mode, which is even suppressed below the phase transition (**Fig. S7**). The characteristic time scales of rotational modes and phonons are compared in **Table 1**.

The complete dynamic picture we provide above allows to understand thermal and charge transports. The product of lifetime and group velocity defines the mean free path, which measures how far phonons travel with carrying heat during the lifetimes[3]. Thus, larger lattice thermal conductivity is expected at longer mean free paths. At 180 K, short lifetimes and small velocities yield mean free paths of ~ 125 and 42 Å for the LA phonon and TA phonon, respectively, which are comparable to those of glasses[27]. Given that the glassy-like thermal conductivity has been previously realized in caged intermetallics through rattling-mode-phonon coupling in the paradigm of electron-crystal phonon-glass[28], the rotor-phonon coupling found here indeed provides an alternative approach to the suppression of thermal transports. It is considerably promising to achieve superior thermoelectric materials in hybrid inorganic-organic perovskites by narrowing their band gaps to enhance charge transports[29]. Meanwhile, the giant electron-phonon coupling ensures phonons strongly interact with carriers through absorption and emission[30]. These short-lived phonons facilitate more frequent scattering of charge carriers, for which charge transports are anticipated to be slowed down. The optical phonons are more relevant due to their closer coupling of rotational modes. In fact, recent photoluminescence experiments suggest that LO phonons through Fröhlich interaction limit the mobility[15].

In summary, we have determined the individual characteristic time scale of various dynamic behaviours in $CH_3NH_3PbI_3$. It is the rotor-phonon coupling that the transports are suppressed. The presence of organic cation characteristic of soft matter is extraordinarily unique because the rotational modes are extremely susceptible to temperature and the existence of hydrogen-bonding enables the stronger linkage to the frame. These fundamental results are beneficial to further study of perovskite solar cells.

**Methods**

Single crystal growth and mounting: The crystal was grown by using the solution method described in ref[10]. The single crystal used in inelastic neutron scattering measurements is 2.32 grams in weight and well-faceted. The as-grown single crystal was sealed into a plastic package under vacuum for storage and transportation. The crystal was mounted onto an aluminium holder in a glove box under helium atmosphere. They were sealed into an aluminium can by using indium wire in the glove box. The whole process was completed without exposure to air in order to minimize the contamination by humidity and oxygen. Photographs are shown in **Fig. S1**.

Inelastic neutron scattering measurements: The high-resolution inelastic neutron scattering measurements were performed at inverse-geometry time-of-flight chopper spectrometer BL02 DNA with $E_i = 2.084$ meV and at direct-geometry time-of-flight chopper spectrometer BL14 AMATERAS with $E_i$ of 54, 16, 7 and 4 meV of J-PARC in Japan[31,32]. The resolutions are summarized in **Table S1**. The crystal was aligned at room temperature at DNA. It took 12 hours to cool down to 140 K from room temperature, avoiding the strain generated during the tetragonal to orthorhombic phase transition. The four-dimension $S(\boldsymbol{q},E)$ data were reduced and visualized by using Utsusemi suite[33], along [$hh$0], [00$l$] and [$k\,\overline{k}$ 0] based on a tetragonal unit cell ($a = 8.80625$ Å, $c = 12.712$ Å). The component along [$k\,\overline{k}$ 0] is limited by setting -0.1 < $k$ < 0.1. The one-dimension spectra were fitted in PAN of DAVE[34]. The EISF was calculated by following $\mathbf{EISF}(\boldsymbol{T}, \boldsymbol{Q}) = \left[\frac{I_{el}(T,Q)}{I(10\,K,Q)/\exp[-U(10K)*Q^2]}\right]/\lambda(\boldsymbol{T})$, where $I_{el}(T,Q)$ is the elastic intensity at temperature $T$, $U$ is thermal factors of hydrogen atoms, $\lambda(T)$ is the inception of linear fitting of log[$I_{el}(T,Q)$] $\sim Q^2$. Note that we used real momentum transfer $Q$ to compare EISF of two directions.

**Acknowledgements:** We acknowledge the award of beam time from J-PARC via proposals, No. 2015A0075 and 2012P0906. Y.L. and S.Z.L. were supported by the National Key Research project

MOST (2016YFA0202400), National Natural Science Foundation of China (61674098), the 111 Project (B14041), and the Chinese National 1000-talent-plan program (1110010341). We thank Dr. R. Kajimoto, Dr. M. Nakamura and Dr. R. Kiyanagi for fruitful discussion.

**Author contribution:** B.L. proposed the project. Y.L. and S.L. synthesized the sample. B.L., M.M., K.S., S.K., T.Y., and K.N. performed the inelastic neutron scattering measurements. B.L. analysed all experimental data. B.L. and S.L. wrote the manuscript with discussion and input from all the authors.

**Additional information:**

**Supplemental Information** accompanies this paper at http://www.nature.com

**Competing financial interests:** The authors declare no competing financial interests.

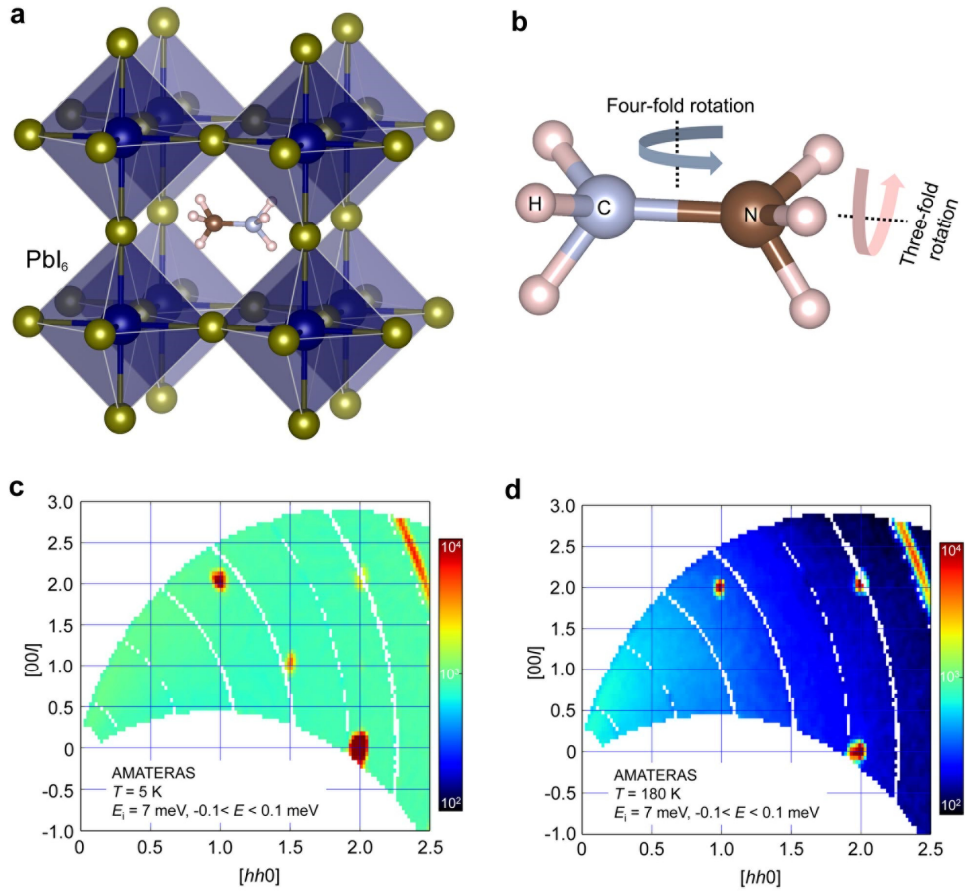

**Figure 1: Crystal structure and phase transition of CH₃NH₃PbI₃. a**. Perovskite structural unit. The organic cation is located in the center of $PbI_6$ octahedral cage. **b**. Geometry of the rotor with schematic drawing of the three-fold rotational mode and four-fold rotational mode. **c** and **d**. The elastic cut (-0.1 < $E$ < 0.1 meV) of $S(\boldsymbol{q},E)$ obtained at AMATERAS displaying the reciprocal-space map along [$hh$0] and [00$l$] directions at 5 and 180 K, where the system is in orthorhombic and tetragonal phases, respectively. Superlattice reflection $\left(\dfrac{3}{2}\dfrac{3}{2}1\right)$ with respect to the tetragonal notion is found at 5 K.

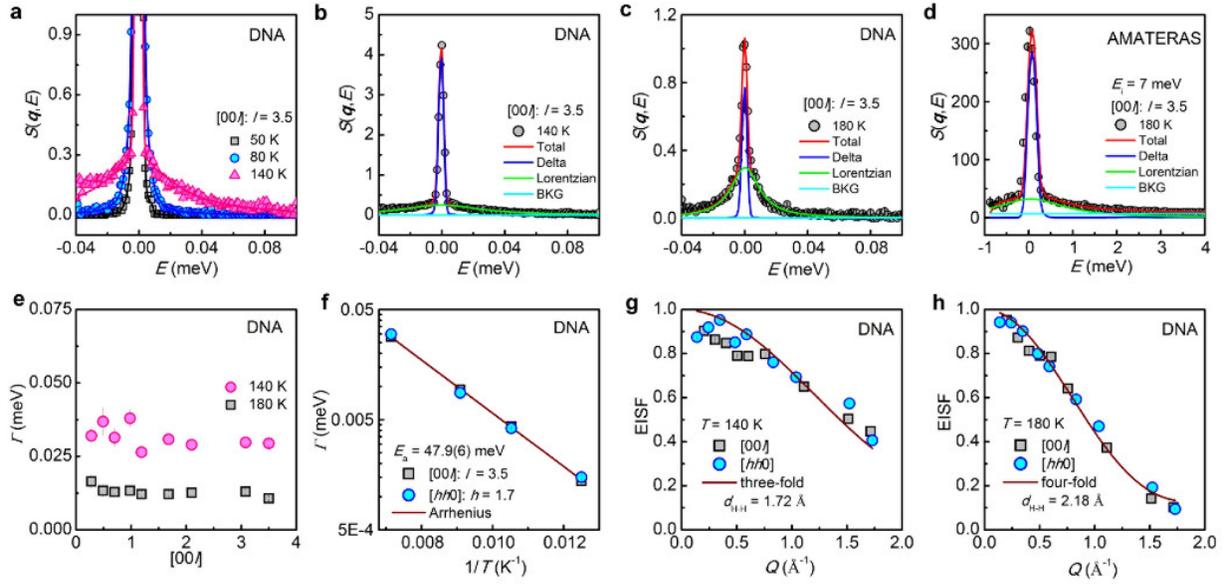

**Figure 2: Rotational dynamics of protons in CH₃NH₃PbI₃. a**. $S(\boldsymbol{q},E)$ at $l$ = 3.5 along [00$l$] at 50, 80, and 140 K obtained at DNA. **b**. The spectrum fitting of $S(\boldsymbol{q},E)$ at 95 K for $l$ = 0.28 and 3.5 along [00$l$] direction, by including a delta function, a Lorentzian function and a constant background (BKG). **c** and **d**. The spectrum fitting of $S(\boldsymbol{q},E)$ obtained at DNA and at AMATERAS at $l$ = 3.5 along [00$l$] direction at 180 K. **e**. The half width at half maximum ($\Gamma$) of the Lorentzian components derived in the fitting, as a function of momentum transfer at 140 and 180 K. **f**. The temperature dependence of $\Gamma$ for orthorhombic phase, fitted to Arrhenius relation $\Gamma \sim \exp(-\dfrac{E_a}{k_B T})$, where $E_a$ is the activation energy of the rotational mode per chemical formula CH₃NH₃PbI₃ and $k_B$ is Boltzmann constant. **g** and **h**. the EISF at 140 and 180 K compared with the three-fold and the four-fold rotational models, respectively. To compare two directions, the real momentum transfer $Q$ is used.

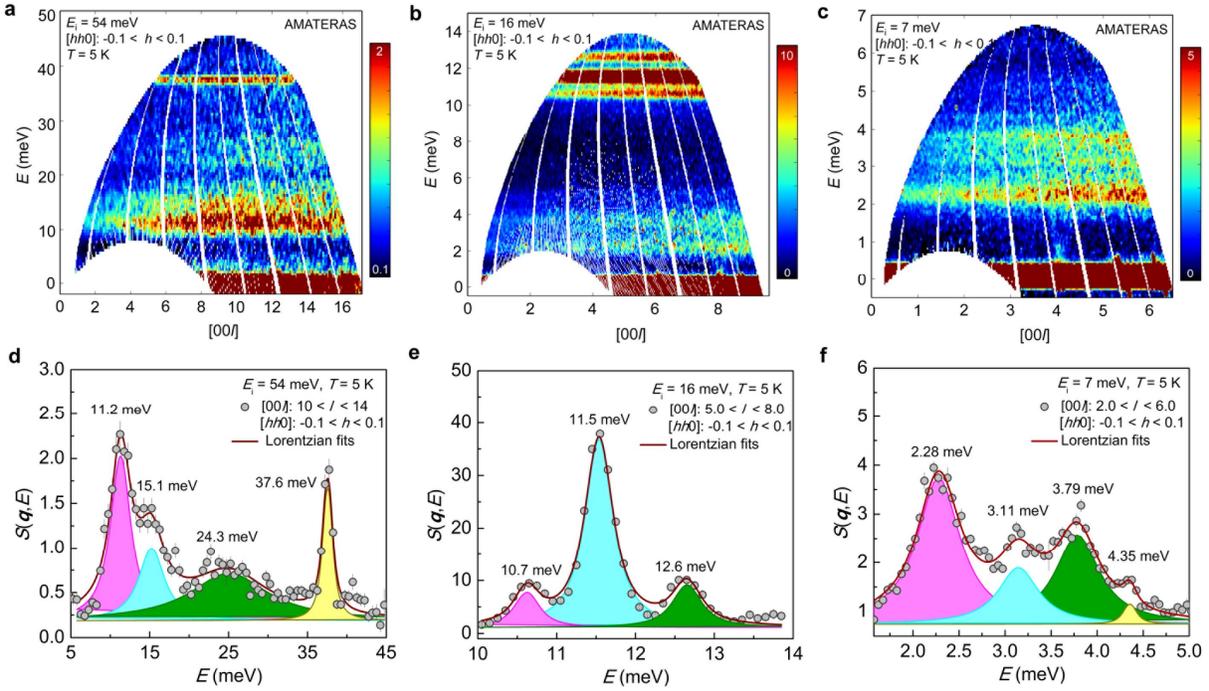

**Figure 3: Longitudinal phonon spectra at 5 K along [00*l*] obtained at AMATERAS. a**, **b** and **c**. The contour plots of $S(\boldsymbol{q},E)$ obtained with $E_i$ of 54, 16, and 7 meV. **d**, **e**, and **f**. Momentum-transfer averaged spectra and multiple Lorentzian-function fitting to highlight the energies of modes. The peak position for each component is labelled. For the case of [*hh*0], please refer to **Fig. S3**.

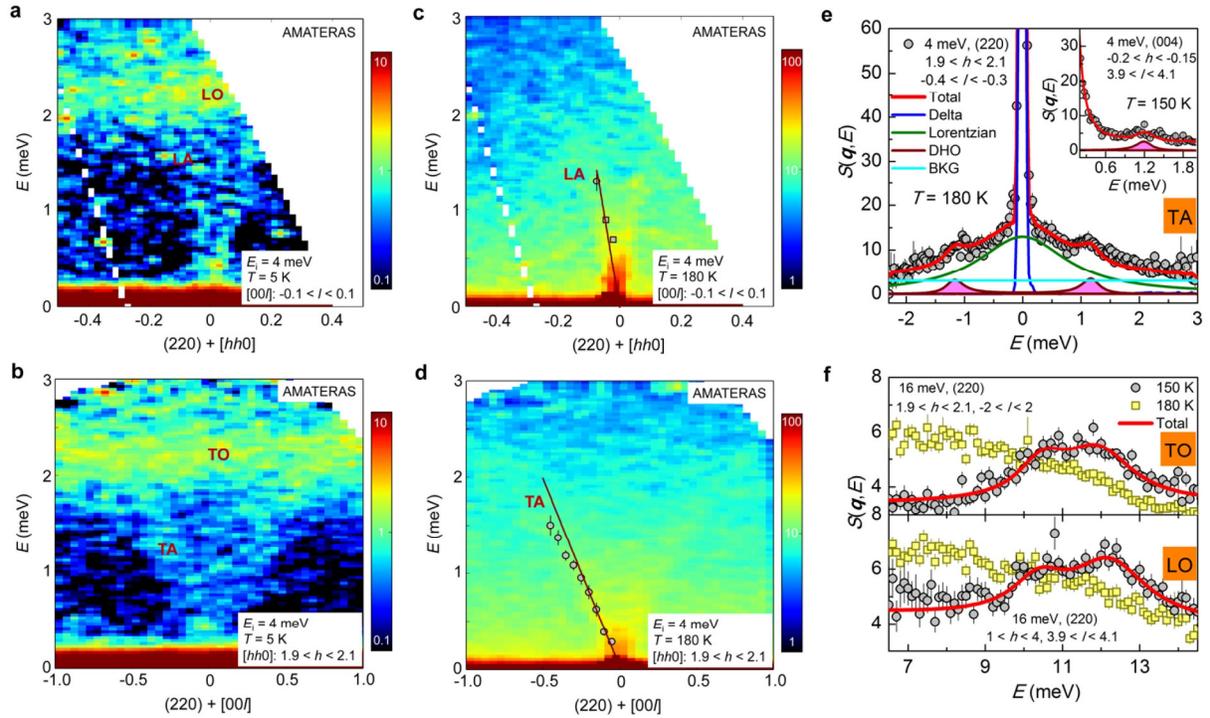

**Figure 4: Influences of proton rotations on phonons. a** and **c**. Longitudinal phonons at 5 and 180 K of the (220) Brillouin zone; **b** and **d**. Transverse phonons at 5 and 180 K of the (220) Brillouin zone. TA, TO, LA and LO phonons are labelled. The filled circles and square in **c** and **d** are dispersions determined by fitting energy-transfer averaged and momentum-transfer averaged spectra, respectively. The solid lines represent the slopes as approaching the zone center, which characterize the group velocities. **e**. Fitting of momentum-transfer averaged TA phonon spectrum of the (220) Brillouin zone at 180 K. The Lorentzian and damped harmonic oscillator (DHO) functions describe the QENS and TA phonon, respectively. The inset shows the TA phonon at 150 K of the (004) Brillouin zone (the contour maps are shown in **Fig. S6**). **f**. Fitting of momentum-transfer averaged TO (upper) and LO (lower) phonons spectra of the (220) Brillouin zone at 150 and 180 K with $E_i$ of 16 meV.

**Table 1.** The relaxation times of rotational modes and lifetimes of TA and TO phonons at orthorhombic and tetragonal phases. The unit is ps.

| Phases | Rotational modes | | Phonons | |
|---|---|---|---|---|
| | $C_3$ | $C_4$ | TA | TO (12.6 meV) |
| **Orthorhombic (150 K[*])** | 23(1) | frozen | 4.53(9) | 0.62(4) |
| **Tetragonal (180 K)** | 0.71(2) | 64(2) | 3.61(42) | << 0.62 |

[*]Relaxation times of rotational modes for orthorhombic phase are taken at 140 K.

# Rotor-phonon coupling in perovskite CH$_3$NH$_3$PbI$_3$: the origin of exceptional transport properties

Bing Li[1], Yukinobu Kawakita[1], Yucheng Liu[2], Masato Matsuura[3], Kaoru Shibata[1], Seiko Kawamura[1], Takeshi Yamada[3], Kenji Nakajima[1] & Shengzhong (Frank) Liu[2,4]

*[1]Japan Proton Accelerator Research Complex, Japan Atomic Energy Agency, Tokai, Ibaraki 319-1195, Japan.*

*[2]Key Laboratory of Applied Surface and Colloid Chemistry, National Ministry of Education, Institute for Advanced Energy Materials, School of Materials Science and Engineering, Shaanxi Normal University, Xi'an 710119, P. R. China.*

*[3]Neutron Science and Technology Center, Comprehensive Research Organization for Science and Society, Tokai, Ibaraki 319-1106, Japan.*

*[4]Dalian Institute of Chemical Physics, Dalian National Laboratory for Clean Energy, Chinese Academy of Sciences, Dalian 116023, P. R. China.*

Correspondence and request for materials should be addressed to B. L. (bing.li@j-parc.jp) or S. L. (liusz@snnu.edu.cn)

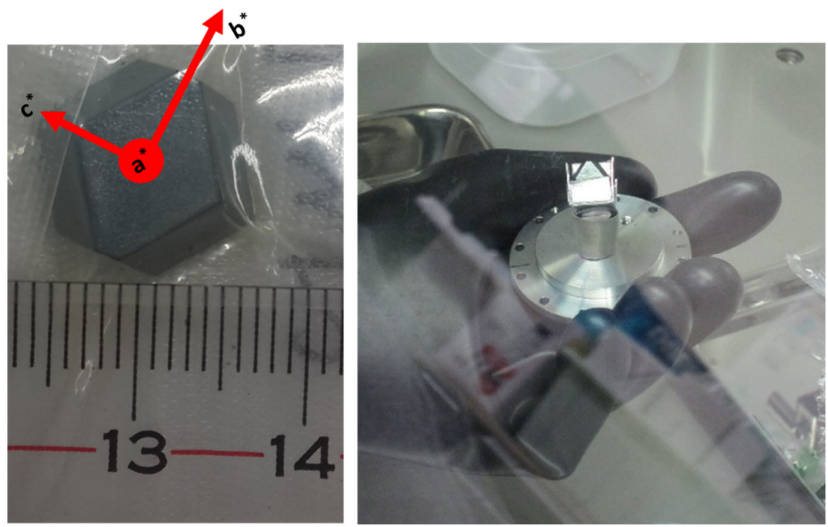

**Figure S1: Photographs of the single crystal CH₃NH₃PbI₃ used in inelastic neutron scattering measurements.** The left panel highlights the dimensions (about 1 cm in length) and the crystallographic directions. The right panel shows the configuration of the crystal attached on the aluminium sample holder.

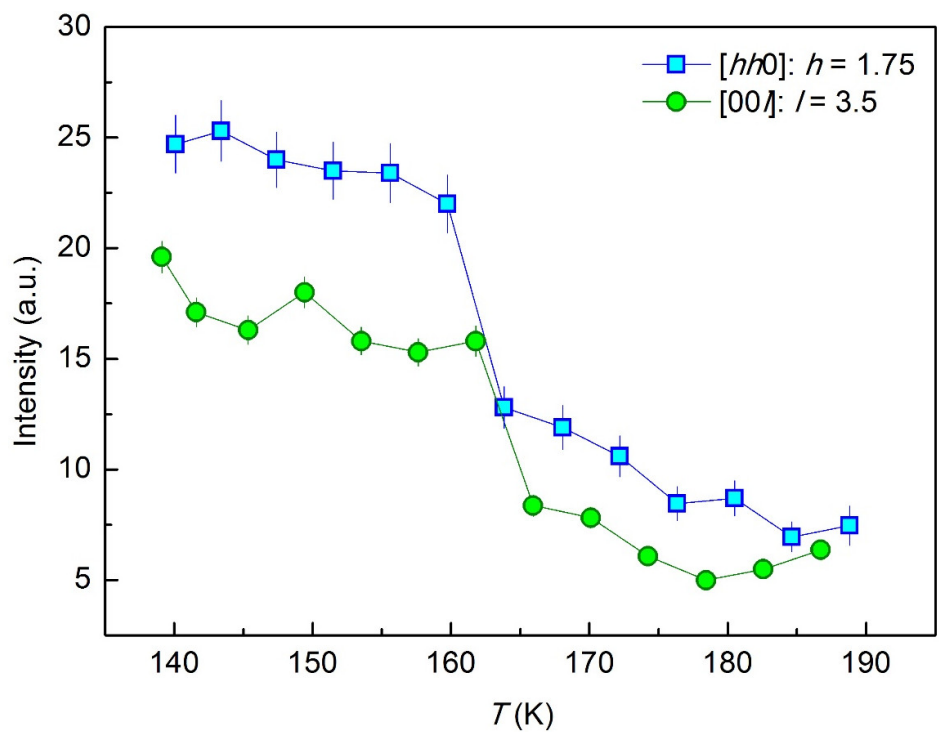

**Figure S2: Temperature dependence of incoherent elastic scattering intensity along two directions.** The drop of intensity around 165 K is due to the orthorhombic to tetragonal phase transition.

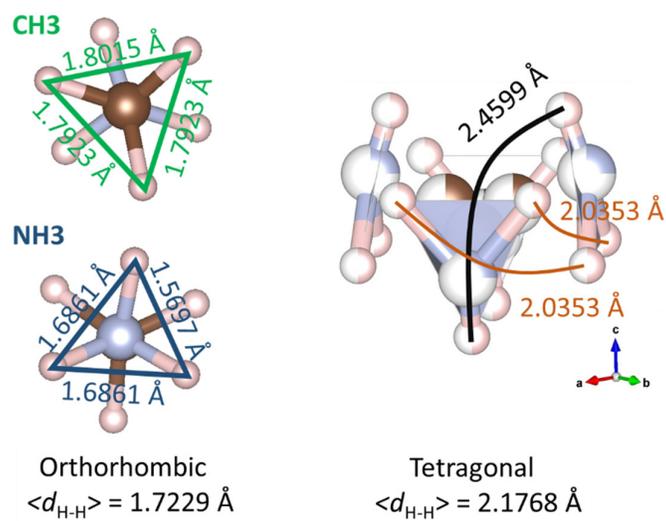

**Figure S3: The H-H bond length in orthorhombic and tetragonal phases.** The crystal structural data is obtained from ref. 20.

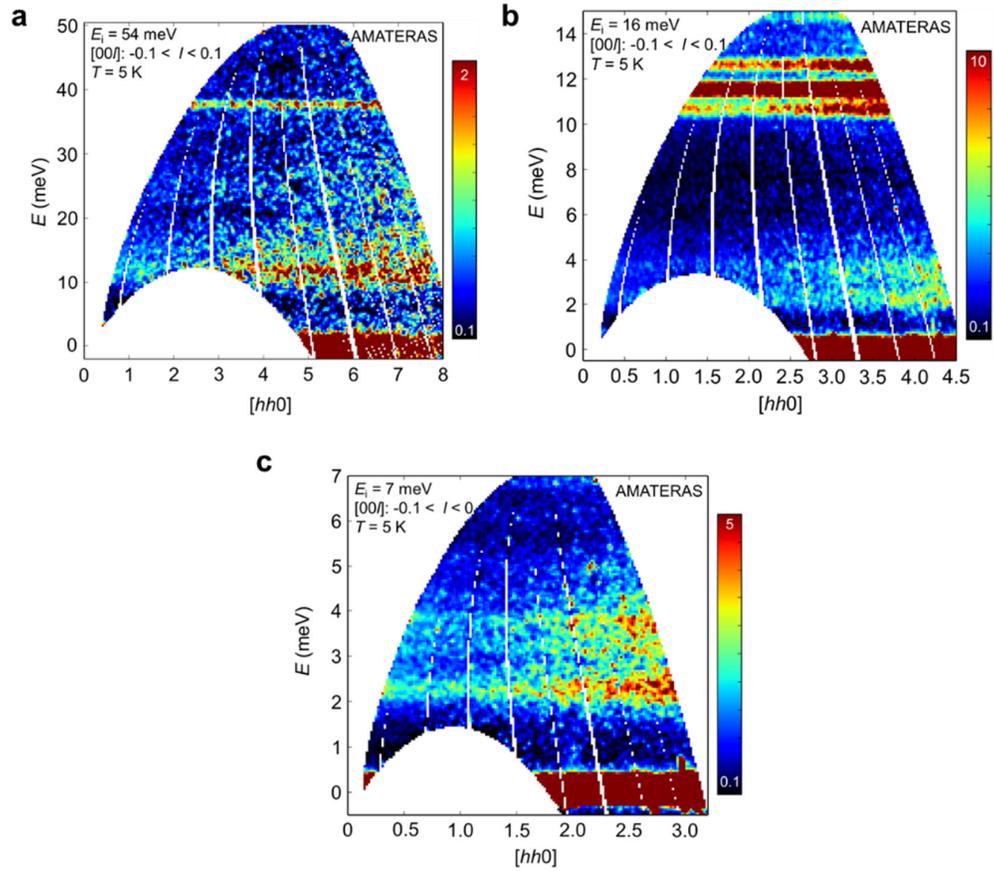

**Figure S4: Longitudinal phonon spectra at 5 K along [*hh*0]. a**, **b** and **c**. The contour plots of $S(\boldsymbol{q},E)$ obtained at AMATERAS with $E_i$ of 54, 16, and 7 meV.

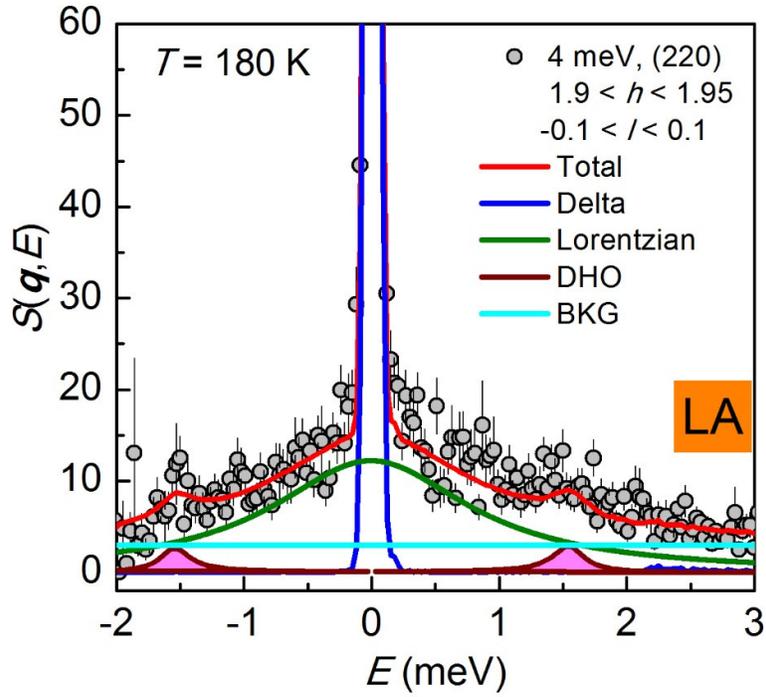

**Figure S5: Longitudinal acoustic (LA) phonon spectrum near the (220) zone center at 180 K.** Multi-component fitting is shown with solid lines, where the delta function carries the elastic component; the Lorentzian function describes the quasi-elastic component originating from rotation modes; damped harmonic oscillator (DHO) function accounts for the LA phonon; and the straight line is for background (BKG).

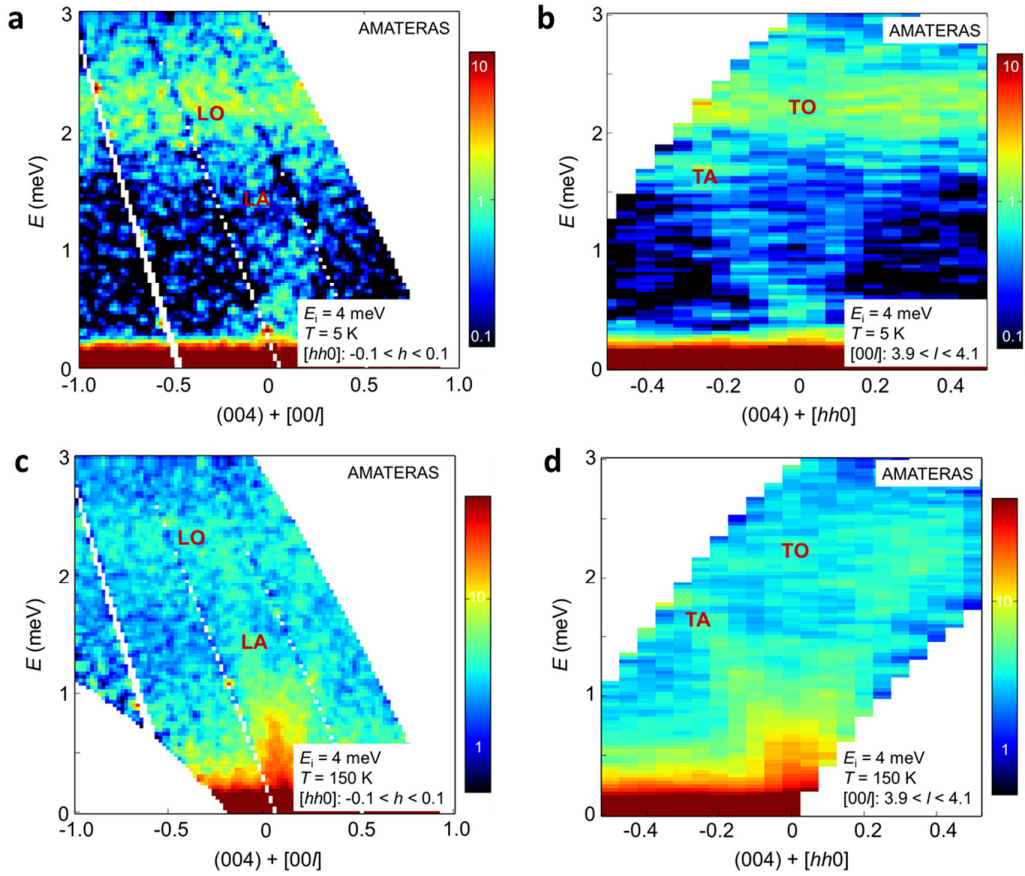

**Figure S6. The contour maps of *S*(*q*,*E*) of the (004) Brillouin zone at 5 and 150 K.** They display longitudinal acoustic (LA), transverse acoustic (TA), longitudinal optical (LO), and transverse optical (TO) phonons, which are labelled. The main information is that optical phonons are present at 150 K and the TA phonon along [*hh*0] is well-defined, in sharp contrast to the data at 180 K shown in **Fig. 4**.

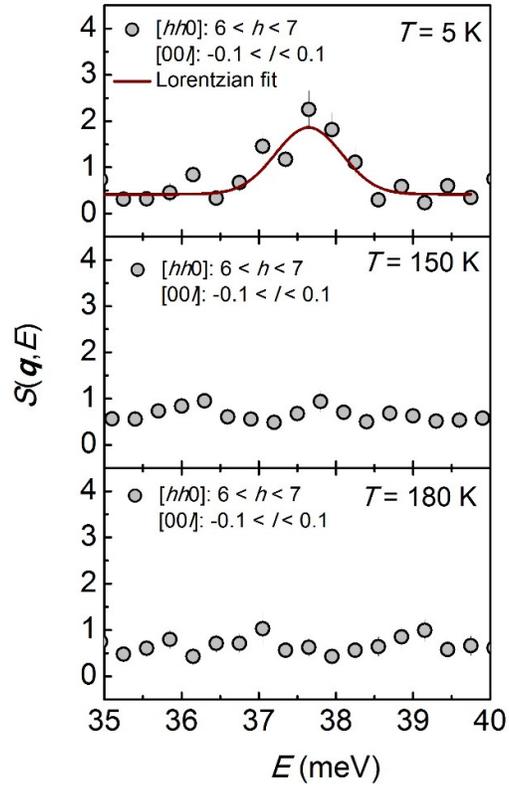

**Figure S7: Influences of proton rotations on the vibrational mode**. The momentum-transfer averaged spectra at 5, 150 and 180 K, with $E_i = 54$ meV. The solid lines show the fitting to Lorentzian function. It can be seen that the vibrational mode at 37.6 meV is completely suppressed at 150 K.

**Table S1. Summary of instrumental resolutions that are determined as the full width at half maximum of the elastic lines.** These resolution functions were used to fit the quasi-elastic scattering component as well as the acoustic phonons. Specially, as for the optical phonons located at about 12 meV, the inelastic resolution function is used, given in brackets.

| Instrument | Resolution (meV) |
|---|---|
| **DNA** | 0.0036 |
| **AMATERAS: $E_i$ = 4 meV** | 0.01 |
| **AMATERAS: $E_i$ = 7 meV** | 0.21 |
| **AMATERAS: $E_i$ = 16 meV** | 0.60 (0.28 at 12 meV) |